%% file: bcjpsipilifetime_v5_proofread.tex
\documentclass[aps,prd,
10pt,
superscriptaddress,
twocolumn,
showpacs
]{revtex4-1}

\usepackage{graphicx}
\usepackage{dcolumn}
\usepackage{bm}
\usepackage{lipsum}

\input definitions
\begin{document}

\title{Measurement of the \bc\ meson lifetime in the decay \bcpsipi}

\input{September2011_Authors}

\date{\today}
\begin{abstract}
The lifetime of the \bc\ meson is measured using 272 exclusive $B_{c}^{-} \to J/\psi (\to \mu^+\mu^-)~\pi^-$ decays 
reconstructed in data 
from proton-antiproton collisions corresponding to an integrated luminosity of
6.7~fb$^{-1}$ recorded by 
the CDF II detector at the Fermilab Tevatron. The lifetime of the \bc\ meson is measured to 
be $\tau$(\bc) = 0.452~$\pm$~0.048(stat)~$\pm$~0.027(syst)~ps. This is the first measurement 
of the \bc\ meson lifetime in a fully-reconstructed hadronic channel, and it agrees with 
previous results and has comparable precision.
\end{abstract}

\maketitle

\newpage

In the standard model the \bc\ meson 
is the only meson composed of two distinct heavy
quarks. The \bc\ meson decay can be governed by the decay of the 
$b$ or $\overline{c}$ spectator quarks or can proceed through the annihilation of the $b$ and $\overline{c}$ quarks. 
Various theoretical techniques have been used to predict the \bc\ meson lifetime. 
An operator-product-expansion calculation~\cite{PhysRevD.53.4991} predicts a \bc\ meson 
lifetime in the range of 0.4 to 0.7~ps. A QCD sum rule approach~\cite{Kiselev2000353} predicts the 
lifetime to be 0.48 $\pm$ 0.05~ps. Another approach~\cite{PhysRevD.64.014003}, estimating the \bc\ meson 
lifetime by global fitting of the phenomenological parameters of all other heavy mesons, gives a result 
of 0.36 or 0.47~ps, depending on different choices of effective heavy-quark masses.

The \bc\ meson lifetime was measured previously in semileptonic decays by CDF~\cite{PhysRevLett.97.012002} 
and D0~\cite{PhysRevLett.102.092001}. 
These measurements have an undetected neutrino in the final state and rely on the modeling of the \bc\ meson
momentum to account for unmeasured momentum of the neutrino.
Therefore, a measurement of the \bc\ meson lifetime in a fully-reconstructed decay mode is 
desired since it does not suffer from this limitation.
CDF is the first experiment to observe the fully-reconstructed $B_{c}^{-} \to J/\psi (\to \mu^+\mu^-)~\pi^-$ decay mode~\cite{conjugate} and to measure  
the \bc\ mass \cite{PhysRevLett.100.182002}. In this paper we present a lifetime measurement of the
\bc\ meson using this decay mode. This measurement is made using data from \ppbar\ collisions at a 
center-of-mass energy of 1.96 TeV recorded by the CDF II detector. The results 
are based on a data sample with an integrated luminosity of 6.7~fb$^{-1}$.


The components of the CDF II detector~\cite{Trigger} most 
relevant for this analysis are the charged-particle tracking and muon identification 
systems. The tracking system is immersed in a uniform 1.4~T solenoidal magnetic field coaxial with the beam 
line, and consists of single- and double-sided silicon detectors \cite{Alan20001} and a 96-layer 
open-cell drift chamber (COT) \cite{Affolder2004249}. 
The muon system is 
used to identify the \psimumu\ decay. Two sets of muon chambers~\cite{CMU, CMX} are used 
to cover different pseudorapidity regions. 

A three-level event-selection system (trigger)  is used  to collect events enriched in \psimumu\ decays~\cite{Trigger}. 
The event reconstruction starts by combining two muon candidates to form a \jpsi\ candidate. The trigger requirements 
are confirmed by selecting events that contain two oppositely charged muon candidates, each with matching COT and muon 
chamber tracks. The muon-pair mass is required to be 
within 80 \mevcc\ of the world-average \jpsi\ mass~\cite{pdg}, where the mass resolution is 13 \mevcc.

Both \psik\ and \psipi\ combinations are reconstructed in this analysis. The large \bupsik\ sample is used 
as a reference decay for \bcpsipi. These final states are identified by assigning the \kaon\ or \pion\ mass to
other reconstructed tracks not used in the \jpsi\ candidates and forming \bu\ or \bc\ candidates. 
Each three-track combination must satisfy a kinematic fit in which the three tracks are required 
to originate from a common decay point, and the invariant mass of the muon pair is constrained to the 
world-average \jpsi\ mass~\cite{pdg}. The \kaon\ and \pion\ candidate track is named the third track $h^-$. 
A minimal selection is made on kinematic quantities after the constrained fit 
including $\pt(h^{-}) > 1.7$~\gevc\ and $\pt(B) > 5$ \gevc, where \pt\ is the momentum component 
transverse to the beam line, and $B$ refers to a $\jpsi~h^-$ candidate. The selection 
criteria for the $B$ candidates are 
listed in Table~\ref{Tab:selection} and discussed below.

\begin{table}[ht]

\caption{Selection requirements.} 
\begin{ruledtabular}
\begin{tabular}{cc}
Selection variable & Requirement  \\ 
\hline
$\pt(h^-)$        & $> 2.0~\gevc$ \\ 
$\pt(B)$  & $> 6.5~\gevc$ \\ 
$P(\chi^{2})$     & $> 0.1\% $ \\ 
$|d(B)|/\sigma_{d(B)}$ & $< 2.0$ \\ 
$\beta_{T}$       & $< 0.2$ radians \\ 
$I_B $            & $> 0.6$ \\
$\sigma_{m}(B)$       & $< 40$~\mevcc \\
$ct$              & $> 80~\mu$m \\
$\sigma_{ct}(B)$      & $< {\rm max}[35, 65-3~\pt(B) (\gevc)]~\mu$m\\
\end{tabular}
\end{ruledtabular}
\label{Tab:selection}
\end{table}

The $h^{-}$ and $B$ candidates are required to have a minimum \pt\ to suppress combinatorial background events. We 
reject events with poorly defined decay points by imposing a lower threshold to the chi-square 
probability $P(\chi^{2})$ of the constrained fit used to reconstruct the $B$ candidates. 
We select $B$ candidates that originate from the primary interaction point by requiring a small impact 
parameter $d(B)$ (transverse distance of closest approach to the beam line) in units of its uncertainty $\sigma_{d(B)}$ and a small angle $\beta_T$ between $\vec{L}_T$ and $\vec{p}_T(B)$, where $\vec{L}_T$ is 
the transverse displacement vector from the primary interaction point to the $B$-decay point, and $\vec{p}_T(B)$ denotes 
a vector in the transverse plane along the momentum direction of the $B$ candidate. The isolation $I_B$ of the $B$ candidate  is 
defined as $I_B~\equiv~p(B)/(p(B)~+~|\sum_{i}\vec{p}_{i}|)$, 
where $\sum_{i}\vec{p}_{i}$ is the sum of momenta over all other reconstructed tracks not used in the $\jpsi~h^-$ combination 
within $\sqrt{(\Delta \eta)^2+(\Delta \phi)^2} < 0.7$, and $\Delta \eta$ and $\Delta \phi$ are the differences in pseudorapidities and 
azimuthal angles of tracks relative to $\vec{p}(B)$.
We also suppress the promptly produced combinatorial background by rejecting candidates with small $ct$, where $ct$ is the  
decay length of the $B$ candidate determined by
 
\begin{equation}
ct~\equiv~\vec{L}_T \cdot \vec{p}_T(B) \frac{c~m(B)}{|p_T(B)|^2},
\label{Eqn:ct}
\end{equation}	
and $m(B)$ is the reconstructed mass of the $B$ candidate. 
Requirements on $\sigma_{m}(B)$ and $\sigma_{ct}(B)$ are 
made to reject poorly reconstructed events, where $\sigma_{m}(B)$ and $\sigma_{ct}(B)$ are the associated uncertainties 
from the kinematic fit of $m(B)$ and $ct(B)$, respectively. The optimization of the selection requirements 
is obtained by maximizing the quantity $\cal S/\sqrt{(S+B)}$ where the background ${\cal B}$ is estimated 
from the mass sidebands in data and the signal ${\cal S}$ is estimated from the signal-region data after
subtracting background contributions as calculated from sidebands.

Because the $\sigma_{ct}(B)$ distribution depends on $\pt(B)$, we vary the requirement on $\sigma_{ct}(B)$ as
a function of $\pt(B)$. 
For candidates 
with $\pt(B)<$~10~\gevc,
we require $\sigma_{ct}(B)~<~(65-3~\pt(B))$~$\mu$m for $\pt(B)$ measured in \gevc, and
$\sigma_{ct}(B)~<~35$~$\mu$m for $\pt(B)\geq$~10~\gevc. 
This \pt-dependent requirement 
on $\sigma_{ct}(B)$ is chosen to be highly efficient for preserving signal while 
reducing combinatorial background and leads to no measurable biases.

\begin{figure}
\centering 
\includegraphics[width=\linewidth]{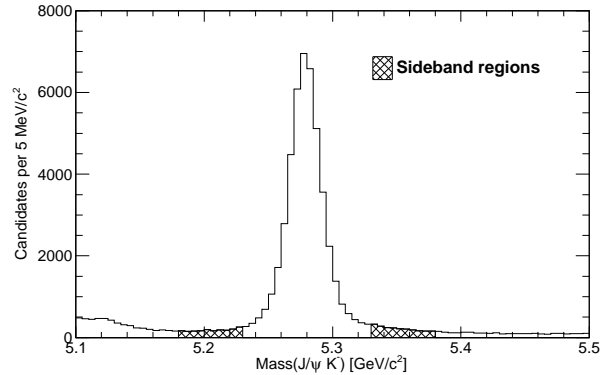} 
\caption{Invariant-mass distribution of $\psik$ candidates. The hatched areas are the 
sideband regions and the signal region lies between them.}
\label{Fig:bupsik_mass}
\end{figure}

The resulting \bu\ mass distribution is shown in Fig.~\ref{Fig:bupsik_mass}. 
The signal region lies between two background sideband regions and has 46 280 \bu\ candidates. The 
two sideband regions consist of a lower sideband from 5.18 to 5.23 \gevcc\ and an upper sideband from 5.33 to 5.38 \gevcc, as shown in the hatched areas.

\begin{figure}
\centering 
\includegraphics[width=\linewidth]{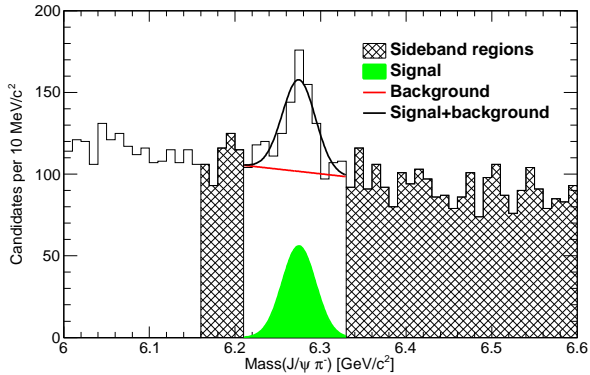} 
\caption{Invariant-mass distribution of $\psipi$ candidates. The hatched areas are the 
sideband regions and the signal region lies between them. The fit result is overlaid in the 
signal region, as well as the signal and background components.}
\label{Fig:bcpsipi_mass}
\end{figure}

The \bcpsipi\ candidates are formed from the same parent sample as
the \bupsik\ candidates where the only change to the reconstruction
is to assign the pion mass to the third track.
We then select events for further analysis using the selections
in Table~\ref{Tab:selection}.
The reconstructed mass distribution for the \bc\ candidates is shown in Fig.~\ref{Fig:bcpsipi_mass}. 
The signal region lies between two background sideband regions and has 1496 \bc\ candidates. The two 
sideband regions of \bc\ candidates consist of a lower sideband from 6.16 to 6.21~\gevcc\ and an upper 
sideband from 6.33 to 6.60~\gevcc, as shown in the hatched areas. The lower sideband is
narrow to avoid contamination from semileptonic \bc\ decays where the lepton is misidentified as a pion.

\begin{figure}
\centering 
\includegraphics[width=\linewidth]{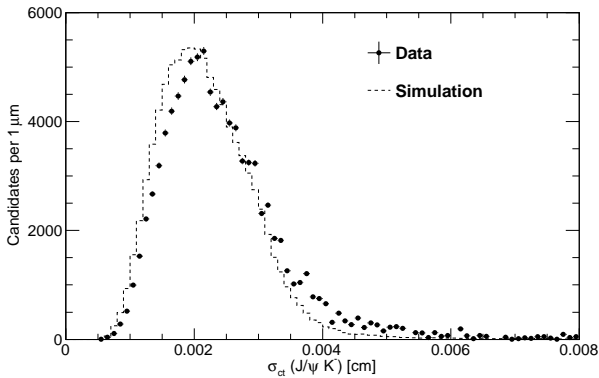} 
\caption{The $\sigma_{ct}$ distribution of $\psik$ candidates obtained from the simulation is compared with data.}
\label{Fig:sigma_ct}
\end{figure}

We generate Monte Carlo simulations for \bupsik\ and \bcpsipi\ decays to study the efficiency 
of the selection criteria as a function of decay length.
The Fixed-Order plus Next-to-Leading 
Logarithms (FONLL) $\pt$ spectrum~\cite{JHEP.07.033} is used for the \bu\ production spectrum. 
We use the calculation of Ref~\cite{PhysRevD.72.114009} as the spectrum for \bc\ simulation.
In comparing the FONLL \pt\ spectrum for \bu\ production with the experimental data, reasonable consistency 
is observed for $\pt>$~6~\gevc, and any residual discrepancy gives a negligible systematic uncertainty. 
To further 
validate the \bupsik\ simulation, we compare 
the distributions of the selection variables listed in Table~\ref{Tab:selection} for experimental 
data and simulation. Generally, good agreement is observed for all selection variables except for $\sigma_{ct}(B)$, whose comparison is shown in Fig.~\ref{Fig:sigma_ct}. 
The disagreement in the $\sigma_{ct}(B)$ distribution arises from mis-modeling of the silicon 
tracking detector in the  simulation, giving a smaller $\sigma_{ct}(B)$ compared with experimental 
data for a given $\pt(B)$.
Consequently, the selection requirement made on $\sigma_{ct}(B)$ for simulation is tuned in order to 
allow the same efficiency as in the experimental data. These $\sigma_{ct}(B)$ selection values for  
simulation are also $\pt$-dependent, and the $\pt$ threshold remains the same; the only change is to 
require $\sigma_{ct}(B)~<~(
45-2~\pt(B))$~$\mu$m for $\pt(B)<$~10~\gevc, and
$\sigma_{ct}(B)~<~25$~$\mu$m for $\pt(B)~\geq~$10~\gevc. The systematic uncertainty associated with the tuning will be  discussed later.

\begin{figure}
\centering 
\vspace{0.5ex}
\includegraphics[width=\linewidth]{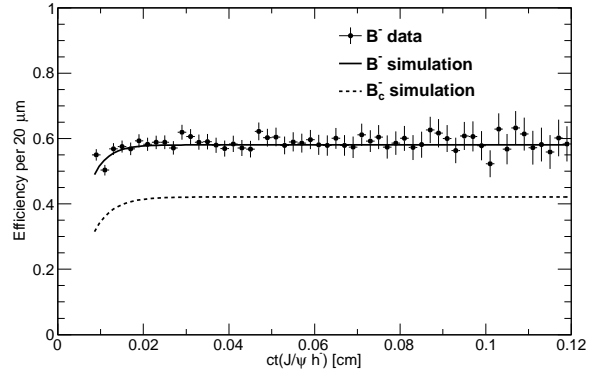} 
\caption{Comparison of efficiency for \bupsik\ obtained from data and the fit result from 
simulation. Also shown is the fit result for \bcpsipi\ simulation.}
\label{Fig:effi}
\end{figure}

The efficiency of the selection criteria is found by comparing the decay-length distribution 
after applying the selection in Table~\ref{Tab:selection} to that obtained from the minimal selection 
which requires $\pt(h^{-}) > 1.7$~\gevc\ and $\pt(B) > 5$ \gevc.
The efficiency determined from  simulation is then fit to a function of the form 

\begin{equation}
\epsilon(ct) = C\left[ 1-\exp\left( \frac{a-ct}{b}\right) \right],
\label{Eqn:effi}
\end{equation}
where $C$, $a$, $b$ are parameters to be fit. 
Figure~\ref{Fig:effi} 
shows the efficiency determined from \bupsik\ experimental data as well as the fit result from  
simulation. The parameter $C$ in Eq.~(\ref{Eqn:effi}) is not necessary in the lifetime fit because 
only the relative shape of the efficiency function matters. 
The requirement on $\beta_{T}$ leads to an efficiency that is not constant as a function of
decay length.
This variable is very effective in rejecting background 
events, especially for events with small $ct$. 
The good agreement between the simulated efficiency and the data-determined efficiency 
supports the use of this approach in the $ct$-dependent efficiency.
The efficiency for the \bcpsipi\ decay as a function of decay time determined from simulation is fit and also shown in Fig.~\ref{Fig:effi}.

We use a maximum log-likelihood simultaneous fit to the 
unbinned  mass and decay-length distributions of the \bc\ candidates. The likelihood 
function consists of signal and background parts, and each part has a mass term and a 
decay length term. The log-likelihood function is
\begin{widetext}
\begin{equation}
\ln {\cal L} =  \sum_{i} \ln\left[ f_s~M_s(m_i)~T_s(ct_i) + (1-f_s)~M_b(m_i)~T_b(ct_i)\right], 
\label{Eqn:likeli}
\end{equation}
\end{widetext}
where $f_s$ is the signal fraction, $m_i$ and $ct_i$ are the reconstructed mass and decay length 
for event $i$. $M_s(m_i)$ and $T_s(ct_i)$ are the normalized probability density functions for mass and 
decay length of the signal model, and $M_b(m_i)$ and $T_b(ct_i)$ are the corresponding functions 
of the background model. The signal mass model $M_s(m_i)$ is described by a Gaussian distribution with 
mean $m_0$ and width $\sigma_{m}$, whose values are determined by the fit. The signal decay length 
model $T_s(ct_i)$ is an exponential distribution with characteristic lifetime $\tau$, smeared by the detector 
resolution and multiplied by the efficiency function given in Eq.~(\ref{Eqn:effi}). The detector 
resolution, which is modeled as a Gaussian distribution centered at zero with a width of 20 $\mu$m, is 
chosen to be consistent with calibration using promptly decaying background events~\cite{PhysRevLett.100.161802}. The background mass 
model $M_b(m_i)$ is described by a linear distribution, and $T_b(ct_i)$ is described by a linear combination of 
three exponential distributions.

A two-step process is used to extract the lifetime of the \bc\ meson. The first step includes the efficiency 
fit and the sideband fit. The efficiency fit is performed on the simulated events using Eq.~(\ref{Eqn:effi}), and 
the result is shown in Fig.~\ref{Fig:effi}. The sideband fit consists of two separate fits in the sideband 
regions. The first fit determines the background mass model parameters, and the second fit determines the background 
decay-time model parameters. 
The signal region is fit with the efficiency and background parameters Gaussian-constrained to the result of the earlier
fits.
The signal fraction and the signal model parameters are allowed to float freely in maximizing the 
log-likelihood function given in Eq.~(\ref{Eqn:likeli}). 

\begin{figure}
\centering 
\includegraphics[width=\linewidth]{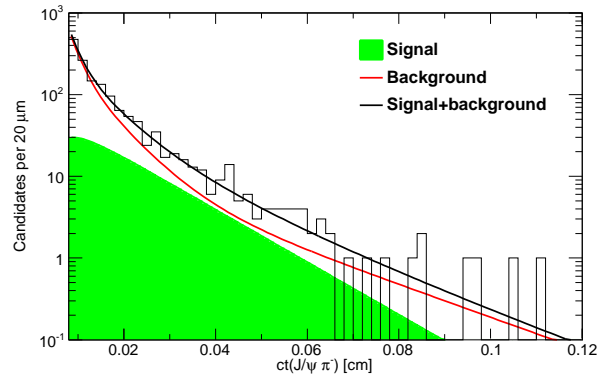} 
\caption{Decay-length distribution of \psipi\ candidates. The fit projection, along individual contributions from signal and background, is
overlaid.}
\label{Fig:sel_ctau}
\end{figure}

To validate this fitting technique, the fit is first applied to the \bu\ candidates shown 
in Fig.~\ref{Fig:bupsik_mass} to extract the \bu\ meson lifetime. The \bu\ lifetime which
is found to be $\tau(\bu)~=~1.637~\pm~0.010~({\rm stat})$~ps is in agreement with the world-average 
value of 1.641 $\pm$ 0.008~ps~\cite{pdg}. The \bu\ signal yield returned from the fit is 43 308 $\pm$ 171.
The fit is then applied to the \bc\ candidates 
shown in Fig.~\ref{Fig:bcpsipi_mass}, resulting in a lifetime of $\tau(\bc)~=~0.452~\pm$~0.048~({\rm stat})~ps for the \bc\ meson, which is 
taken as our central result, with a \bc\ signal yield of 272 $\pm$ 61~({\rm stat}) candidates.  The \bc\ meson mass of 6274.6 $\pm$ 2.9~({\rm stat}) \mevcc\ returned from 
the fit is in good agreement with the previous CDF determination~\cite{PhysRevLett.100.182002}. 
Figures~\ref{Fig:bcpsipi_mass} and~\ref{Fig:sel_ctau} 
show the distribution of the \psipi\ mass and decay length. The fit projections are overlaid.

We have considered several sources of systematic uncertainty and evaluated 
their contributions. 
To evaluate possible systematic uncertainties with the models in the likelihood function, we generate
400 simulated data samples (pseudoexperiments) whose distributions are
based on the fit results determined by the experimental data. These pseudoexperiments are then fit with 
the default and alternate models separately. The distributions of the sample-by-sample lifetime differences 
between different models are obtained and compared with the differences observed in experimental data.
To assess the effect of the choice
of the linear model for the mass-fit background, we compare to the result of a fit using
a bilinear model that allows the background distribution to 
have different slopes at masses lower and higher than the \bc\ pole mass, with the constraint that these 
two distributions intersect at the fit \bc\ mass value. The fit lifetime with this bilinear 
model has a shift of --0.009~ps compared with the default linear model. The pseudoexperiments suggests
up to a 0.017~ps difference from this variation. We
conclude that the shift between the data fits is
consistent with the spread among the pseudoexperiments, and we use
that larger difference as the systematic uncertainty from the
background mass model.

To assess a possible systematic uncertainty due to the modeling of the long tail in the background 
decay-length distribution, we test an alternate model of the background decay time 
which uses a linear distribution to replace the component with the largest characteristic 
lifetime in the three exponential distributions. This variation gives a lifetime result that 
changes by  \mbox{--0.0007~ps} compared with the default background decay-time model. However, fit 
results from pseudoexperiments suggest the difference between these two models could 
be 0.013~ps, which is included as the systematic uncertainty due to the choice of the background 
decay-time model.

The signal decay-time model includes the efficiency determined from the simulation. 
We have performed several studies to estimate the associated systematic uncertainty. 
First, the fit is repeated using an efficiency
function obtained without  tuning the $\sigma_{ct}(B)$ difference between data and simulation. 
The difference in the estimated lifetime is 0.003~ps. Second, the 
efficiency function is shifted toward lower and higher decay length by 20 $\mu$m to account 
for a possible uncertainty in determining the efficiency function parameters; this 20 $\mu$m shift is 
equivalent to three standard deviations of the parameter $a$ in Eq.~(\ref{Eqn:effi}). This variation 
gives a difference of --0.010 (+0.007)~ps for shifting toward lower (higher) decay lengths. 
The distribution of the difference between the resulting 
lifetimes in the pseudoexperiments is fit by a 
Gaussian distribution that centers at --0.006 (0.004)~ps with a width of 0.002 (0.001)~ps for shifting 
toward lower (higher) decay lengths.  
Third, the systematic uncertainty associated with the \bc\ production spectrum has been 
assessed. We vary the relative fraction of different contributions to the production spectrum; the 
difference in the corresponding efficiency is negligible and no systematic uncertainty is assigned to it. 
Finally, to further study the systematic uncertainty associated with the production spectrum, we use 
the efficiency parameters obtained from the \bupsik\  simulation. Since the \bu\ production spectrum 
is quite different from that of \bc, the fit lifetime difference of 0.007~ps indicates that the production 
spectrum does not contribute significantly to the systematic uncertainty. 
Thus, the total systematic uncertainty associated with the signal decay-time model is taken to be 0.010~ps.

Correlations between the lifetime and other parameters of the analysis are considered as 
possible systematic uncertainties. The list of parameters includes the minimum 
and maximum decay length for events in the final fit, adding a parameter 
to the efficiency model, small variations in the sideband definitions, small modifications 
in the selection requirements, the use of an alternate fit procedure which fits the sideband 
and the signal regions simultaneously, the mass resolution in the signal model, the background 
fraction, and the three terms describing the exponentials in the background decay time model. 
No systematic effect was found to significantly exceed the variations expected from statistical 
uncertainties. We assign an additional uncertainty of 0.010~ps as a conservative approach to 
account for possible small systematic effects.

The systematic uncertainty due to tracking detector misalignments is evaluated by generating 
simulated samples with radial displacements of individual sensors as well as translation and 
rotation of the silicon detector relative to the COT~\cite{PhysRevLett.106.121804}. A systematic 
uncertainty of 0.007~ps is assigned to the misalignment based on these simulated samples. The 
systematic uncertainty from the signal mass model is evaluated by including a contribution 
to the total \bc\ signal yield from the Cabibbo-suppressed decay \bcpsik\ in the signal mass shape.
The Cabibbo-suppressed contribution is fixed to be 5\% of the total \bc\ signal yield as 
determined from the Cabibbo angle. This effect results in a 0.003~ps variation. 

The systematic uncertainty from the fitting technique itself is tested by generating 
pseudoexperiments, and comparing the fit lifetimes with the input lifetime. The bias
on the lifetime returned by the fit is found to be no greater than 0.003~ps which we take
as systematic uncertainty. Table~\ref{Tab:sys} 
summarizes the systematic uncertainties, which are added in quadrature to determine the total systematic uncertainty.

\begin{table}[ht]
\caption{Summary of systematic uncertainties.}
\label{Tab:sys}
\begin{ruledtabular}
\begin{tabular}{lc}
Source & Uncertainty [ps]  \\
\hline
Background mass model        & 0.017 \\
Background decay-time model  & 0.013 \\
Signal decay-time model      & 0.010 \\
Correlation                  & 0.010 \\
Misalignment                  & 0.007 \\
Signal mass model            & 0.003 \\
Fitting technique            & 0.003 \\ \hline
Total                        & 0.027  \\ 
\end{tabular}
\end{ruledtabular}
\end{table}

Given that the efficiency is not uniform over decay length, our result 
relies on the accuracy of the simulation in determining the efficiency. 
We check our result by measuring the \bc\ lifetime using a different set of selection 
criteria, each of which has uniform efficiency in the \bupsik\ decay. The 
most important differences between these selection 
criteria and those listed in Table~\ref{Tab:selection} are removing the $\beta_{T}$ requirement
and using a larger minimum $ct$ requirement. The alternate selection criteria gives 6538 \bc\ candidates between 6.0 
and 6.6 \gevcc\ which is roughly the same number as obtained from the selections in 
Table~\ref{Tab:selection} (6368 candidates), while only 2578 candidates are common to both samples.

The consistency check also uses an unbinned maximum log-likelihood fit to extract 
the \bc\ meson lifetime. 
The signal mass model consists of a Gaussian 
distribution centered at the \bc\ meson mass and a Cabibbo-suppressed \bcpsik\ Gaussian 
distribution centered at 60 \mevcc\ below the \bc\ meson mass with a 30 \mevcc\ width. 
The 
signal decay-time model is an exponential distribution convoluted with the detector 
resolution. The background mass model is described by a linear distribution, and the background 
decay-time model consists of two prompt Gaussian distributions, two positive exponential 
distributions, and one negative exponential distribution.

A similar two-step fit is used in the consistency check. The first step is to determine the background 
parameters from the sideband fit, where the sideband is defined as the \psipi\ invariant-mass region 
between 6.4 and 6.5 \gevcc. The sideband fit is performed on events with decay length 
between --1000 $\mu$m and 1000 $\mu$m and the resulting background parameters are fixed in the second 
step. In the second step we fit events in the signal region between 6.16 
and 6.36 \gevcc, and only the signal fraction and signal model parameters are allowed to float. The 
consistency check is first performed on the \bu\ candidates. The fit result finds the \bu\ lifetime 
to be $\tau=~1.647~\pm$~0.020~({\rm stat})~ps, which agrees with the world-average value of 1.641~$\pm$~0.008~ps~\cite{pdg}. 
The consistency check is then applied to the \bc\ candidates, giving a \bc\ meson lifetime 
of $\tau~=~0.450~\pm$~0.053~({\rm stat})~ps which is consistent with our central value of 0.452~$\pm$~0.048~({\rm stat})~ps. 
The \bc\ signal yield from the consistency check is 308~$\pm$~39~({\rm stat}) which is compared with 272 $\pm$ 61~({\rm stat}) 
from the central result. The total systematic uncertainty in the consistency check is 0.033~ps where 
the largest uncertainty of 0.027~ps comes from the background decay-time model. Thus, we conclude that 
our central result obtained from the $ct$-dependent efficiency is reliable.

In conclusion, we have made the first measurement of the \bc\ meson lifetime in a fully-reconstructed hadronic
decay mode. Using the \bcpsipi\ decay channel, the lifetime of the \bc\ meson is measured to be
$\tau =$ 0.452~$\pm$~0.048(stat)~$\pm$~0.027(syst)~ps. This result is consistent with the most
recent result from the D0 collaboration~\cite{PhysRevLett.102.092001} using 
semileptonic decay channels, $\tau =$ 0.448~$^{+0.038}_{-0.036}$(stat)~$\pm$~0.032(syst)~ps,
and has comparable precision. The result also agrees with theoretical calculations in which
the decay width of the \bc\ meson is dominated by the decay of the charm quark.

We thank the Fermilab staff and the technical staffs of the participating institutions for their vital 
contributions. This work was supported by the U.S. Department of Energy and National Science 
Foundation; the Italian Istituto Nazionale di Fisica Nucleare; the 
Ministry of Education, Culture, Sports, Science and Technology of Japan; the Natural Sciences and 
Engineering Research Council of Canada; the National Science Council of the Republic of China; the Swiss 
National Science Foundation; the A.P. Sloan Foundation; the Bundesministerium f\"ur Bildung und 
Forschung, Germany; the Korean World Class University Program, the National Research Foundation of 
Korea; the Science and Technology Facilities Council and the Royal Society, UK; the Russian Foundation 
for Basic Research; the Ministerio de Ciencia e Innovaci\'{o}n, and Programa Consolider-Ingenio 
2010, Spain; the Slovak R\&D Agency; the Academy of Finland; and the Australian Research Council (ARC).

\clearpage

\end{document}

%% file: definitions.tex
\newcommand{\bu}{\mbox{$B^{-}$}}

\newcommand{\bc}{\mbox{$B_c^{-}$}}

\newcommand{\bupsik}{\mbox{$B^{-} \rightarrow  J/\psi~K^{-}$}}

\newcommand{\bcpsipi}{\mbox{$B_{c}^{-} \rightarrow J/\psi~\pi^{-}$}}
\newcommand{\bcpsik}{\mbox{$B_{c}^{-} \rightarrow J/\psi~K^{-}$}}

\newcommand{\psimumu}{\mbox{$J/\psi \rightarrow \mu^{+}~\mu^{-}$}}

\newcommand{\psipi}{\mbox{$J/\psi~\pi^{-}$}}
\newcommand{\psik}{\mbox{$J/\psi~K^{-}$}}

\newcommand{\ppbar}{\mbox{$p\overline{p}$}}
\newcommand{\pt}{\mbox{$p_{T}$}}

\newcommand{\mevcc}{\mbox{MeV/$c^2$}}
\newcommand{\gevcc}{\mbox{GeV/$c^2$}}

\newcommand{\gevc}{\mbox{GeV/$c$}}
\newcommand{\jpsi}{\mbox{$J/\psi$}}

\newcommand{\pion}{\mbox{$\pi^{-}$}}
\newcommand{\kaon}{\mbox{$K^{-}$}}
\def\lessim{\mathrel {\vcenter {\baselineskip 0pt \kern 0pt
\hbox{$<$} \kern 0pt \hbox{$\sim$}}}}

%% file: September2011_Authors.tex
\affiliation{Institute of Physics, Academia Sinica, Taipei, Taiwan 11529, Republic of China}
\affiliation{Argonne National Laboratory, Argonne, Illinois 60439, USA}
\affiliation{University of Athens, 157 71 Athens, Greece}
\affiliation{Institut de Fisica d'Altes Energies, ICREA, Universitat Autonoma de Barcelona, E-08193, Bellaterra (Barcelona), Spain}
\affiliation{Baylor University, Waco, Texas 76798, USA}
\affiliation{Istituto Nazionale di Fisica Nucleare Bologna, $^{ee}$University of Bologna, I-40127 Bologna, Italy}
\affiliation{University of California, Davis, Davis, California 95616, USA}
\affiliation{University of California, Los Angeles, Los Angeles, California 90024, USA}
\affiliation{Instituto de Fisica de Cantabria, CSIC-University of Cantabria, 39005 Santander, Spain}
\affiliation{Carnegie Mellon University, Pittsburgh, Pennsylvania 15213, USA}
\affiliation{Enrico Fermi Institute, University of Chicago, Chicago, Illinois 60637, USA}
\affiliation{Comenius University, 842 48 Bratislava, Slovakia; Institute of Experimental Physics, 040 01 Kosice, Slovakia}
\affiliation{Joint Institute for Nuclear Research, RU-141980 Dubna, Russia}
\affiliation{Duke University, Durham, North Carolina 27708, USA}
\affiliation{Fermi National Accelerator Laboratory, Batavia, Illinois 60510, USA}
\affiliation{University of Florida, Gainesville, Florida 32611, USA}
\affiliation{Laboratori Nazionali di Frascati, Istituto Nazionale di Fisica Nucleare, I-00044 Frascati, Italy}
\affiliation{University of Geneva, CH-1211 Geneva 4, Switzerland}
\affiliation{Glasgow University, Glasgow G12 8QQ, United Kingdom}
\affiliation{Harvard University, Cambridge, Massachusetts 02138, USA}
\affiliation{Division of High Energy Physics, Department of Physics, University of Helsinki and Helsinki Institute of Physics, FIN-00014, Helsinki, Finland}
\affiliation{University of Illinois, Urbana, Illinois 61801, USA}
\affiliation{The Johns Hopkins University, Baltimore, Maryland 21218, USA}
\affiliation{Institut f\"{u}r Experimentelle Kernphysik, Karlsruhe Institute of Technology, D-76131 Karlsruhe, Germany}
\affiliation{Center for High Energy Physics: Kyungpook National University, Daegu 702-701, Korea; Seoul National University, Seoul 151-742, Korea; Sungkyunkwan University, Suwon 440-746, Korea; Korea Institute of Science and Technology Information, Daejeon 305-806, Korea; Chonnam National University, Gwangju 500-757, Korea; Chonbuk National University, Jeonju 561-756, Korea}
\affiliation{Ernest Orlando Lawrence Berkeley National Laboratory, Berkeley, California 94720, USA}
\affiliation{University of Liverpool, Liverpool L69 7ZE, United Kingdom}
\affiliation{University College London, London WC1E 6BT, United Kingdom}
\affiliation{Centro de Investigaciones Energeticas Medioambientales y Tecnologicas, E-28040 Madrid, Spain}
\affiliation{Massachusetts Institute of Technology, Cambridge, Massachusetts 02139, USA}
\affiliation{Institute of Particle Physics: McGill University, Montr\'{e}al, Qu\'{e}bec, Canada H3A~2T8; Simon Fraser University, Burnaby, British Columbia, Canada V5A~1S6; University of Toronto, Toronto, Ontario, Canada M5S~1A7; and TRIUMF, Vancouver, British Columbia, Canada V6T~2A3}
\affiliation{University of Michigan, Ann Arbor, Michigan 48109, USA}
\affiliation{Michigan State University, East Lansing, Michigan 48824, USA}
\affiliation{Institution for Theoretical and Experimental Physics, ITEP, Moscow 117259, Russia}
\affiliation{University of New Mexico, Albuquerque, New Mexico 87131, USA}
\affiliation{The Ohio State University, Columbus, Ohio 43210, USA}
\affiliation{Okayama University, Okayama 700-8530, Japan}
\affiliation{Osaka City University, Osaka 588, Japan}
\affiliation{University of Oxford, Oxford OX1 3RH, United Kingdom}
\affiliation{Istituto Nazionale di Fisica Nucleare, Sezione di Padova-Trento, $^{ff}$University of Padova, I-35131 Padova, Italy}
\affiliation{University of Pennsylvania, Philadelphia, Pennsylvania 19104, USA}
\affiliation{Istituto Nazionale di Fisica Nucleare Pisa, $^{gg}$University of Pisa, $^{hh}$University of Siena and $^{ii}$Scuola Normale Superiore, I-56127 Pisa, Italy}
\affiliation{University of Pittsburgh, Pittsburgh, Pennsylvania 15260, USA}
\affiliation{Purdue University, West Lafayette, Indiana 47907, USA}
\affiliation{University of Rochester, Rochester, New York 14627, USA}
\affiliation{The Rockefeller University, New York, New York 10065, USA}
\affiliation{Istituto Nazionale di Fisica Nucleare, Sezione di Roma 1, $^{jj}$Sapienza Universit\`{a} di Roma, I-00185 Roma, Italy}
\affiliation{Rutgers University, Piscataway, New Jersey 08855, USA}
\affiliation{Texas A\&M University, College Station, Texas 77843, USA}
\affiliation{Istituto Nazionale di Fisica Nucleare Trieste/Udine, I-34100 Trieste, $^{kk}$University of Udine, I-33100 Udine, Italy}
\affiliation{University of Tsukuba, Tsukuba, Ibaraki 305, Japan}
\affiliation{Tufts University, Medford, Massachusetts 02155, USA}
\affiliation{University of Virginia, Charlottesville, Virginia 22906, USA}
\affiliation{Waseda University, Tokyo 169, Japan}
\affiliation{Wayne State University, Detroit, Michigan 48201, USA}
\affiliation{University of Wisconsin, Madison, Wisconsin 53706, USA}
\affiliation{Yale University, New Haven, Connecticut 06520, USA}

\author{T.~Aaltonen}
\affiliation{Division of High Energy Physics, Department of Physics, University of Helsinki and Helsinki Institute of Physics, FIN-00014, Helsinki, Finland}
\author{B.~\'{A}lvarez~Gonz\'{a}lez$^z$}
\affiliation{Instituto de Fisica de Cantabria, CSIC-University of Cantabria, 39005 Santander, Spain}
\author{S.~Amerio}
\affiliation{Istituto Nazionale di Fisica Nucleare, Sezione di Padova-Trento, $^{ff}$University of Padova, I-35131 Padova, Italy}
\author{D.~Amidei}
\affiliation{University of Michigan, Ann Arbor, Michigan 48109, USA}
\author{A.~Anastassov$^x$}
\affiliation{Fermi National Accelerator Laboratory, Batavia, Illinois 60510, USA}
\author{A.~Annovi}
\affiliation{Laboratori Nazionali di Frascati, Istituto Nazionale di Fisica Nucleare, I-00044 Frascati, Italy}
\author{J.~Antos}
\affiliation{Comenius University, 842 48 Bratislava, Slovakia; Institute of Experimental Physics, 040 01 Kosice, Slovakia}
\author{G.~Apollinari}
\affiliation{Fermi National Accelerator Laboratory, Batavia, Illinois 60510, USA}
\author{J.A.~Appel}
\affiliation{Fermi National Accelerator Laboratory, Batavia, Illinois 60510, USA}
\author{T.~Arisawa}
\affiliation{Waseda University, Tokyo 169, Japan}
% ...

\collaboration{CDF Collaboration\footnote{With visitors from
$^a$Istituto Nazionale di Fisica Nucleare, Sezione di Cagliari, 09042 Monserrato (Cagliari), Italy,
$^b$University of CA Irvine, Irvine, CA 92697, USA,
$^c$University of CA Santa Barbara, Santa Barbara, CA 93106, USA,
$^d$University of CA Santa Cruz, Santa Cruz, CA 95064, USA,
$^e$Institute of Physics, Academy of Sciences of the Czech Republic, Czech Republic,
$^f$CERN, CH-1211 Geneva, Switzerland,
$^g$Cornell University, Ithaca, NY 14853, USA,
$^h$University of Cyprus, Nicosia CY-1678, Cyprus,
$^i$Office of Science, U.S. Department of Energy, Washington, DC 20585, USA,
$^j$University College Dublin, Dublin 4, Ireland,
$^k$ETH, 8092 Zurich, Switzerland,
$^l$University of Fukui, Fukui City, Fukui Prefecture, Japan 910-0017,
$^m$Universidad Iberoamericana, Mexico D.F., Mexico,
$^n$University of Iowa, Iowa City, IA 52242, USA,
$^o$Kinki University, Higashi-Osaka City, Japan 577-8502,
$^p$Kansas State University, Manhattan, KS 66506, USA,
$^q$Korea University, Seoul, 136-713, Korea,
$^r$University of Manchester, Manchester M13 9PL, United Kingdom,
$^s$Queen Mary, University of London, London, E1 4NS, United Kingdom,
$^t$University of Melbourne, Victoria 3010, Australia,
$^u$Muons, Inc., Batavia, IL 60510, USA,
$^v$Nagasaki Institute of Applied Science, Nagasaki, Japan,
$^w$National Research Nuclear University, Moscow, Russia,
$^x$Northwestern University, Evanston, IL 60208, USA,
$^y$University of Notre Dame, Notre Dame, IN 46556, USA,
$^z$Universidad de Oviedo, E-33007 Oviedo, Spain,
$^{aa}$CNRS-IN2P3, Paris, F-75205 France,
$^{bb}$Texas Tech University, Lubbock, TX 79609, USA,
$^{cc}$Universidad Tecnica Federico Santa Maria, 110v Valparaiso, Chile,
$^{dd}$Yarmouk University, Irbid 211-63, Jordan,
}}
\noaffiliation